\documentclass{aastex62}

\newcommand\be{\begin{equation}}
\newcommand\ee{\end{equation}}
\usepackage{rotating}

\submitjournal{ApJ}

\shorttitle{Magnetar origin of X-ray afterglow GRBs}
\shortauthors{Stratta, Dainotti et al.}

\begin{document}

\title{On the magnetar origin of the GRBs presenting X-ray afterglow plateaus}

\correspondingauthor{Maria Giovanna Dainotti\footnote{The first and the second author contributed equally}}
\email{mdainott@stanford.edu}

\author[0000-0002-0786-7307]{Giulia Stratta}
\affiliation{Urbino University, via Santa Chiara 27, 61027 \\
Urbino, Italy}
\affiliation{INAF OAS  Osservatorio di Astrofisica e Scienza dello Spazio di Bologna \\
Plesso del Battiferro, Via Gobetti 93/3, 40129 Bologna, Italy}

\author{Maria Giovanna Dainotti}
\affiliation{Department of Physics \& Astronomy Stanford University\\
Via Pueblo Mall 382, Stanford CA, 94305-4060, USA}
\affiliation{Obserwatorium Astronomiczne, Uniwersytet Jagiellonski\\ 
ul. Orla 171, 31-501 Krak\'ow, Poland}
\affiliation{INAF OAS  Osservatorio di Astrofisica e Scienza dello Spazio di Bologna \\
Plesso del Battiferro, Via Gobetti 93/3, 40129 Bologna, Italy}

\author{Simone Dall'Osso}
\affiliation{Dept. of Physics \& Astronomy, Stony Brook University \\
Stony Brook, NY 11794-3800, USA}

\author{X. Hernandez}
\affiliation{Instituto de Astronom\'{\i}a, Universidad Nacional Aut\'{o}noma
  de M\'{e}xico, \\Apartado Postal 70--264 C.P. 04510 M\'exico D.F. M\'exico}
	
	\author{Giovanni De Cesare}
\affiliation{INAF OAS  Osservatorio di Astrofisica e Scienza dello Spazio di Bologna \\
Plesso del Battiferro, Via Gobetti 93/3, 40129 Bologna, Italy}

\begin{abstract}
The X-ray afterglow plateau emission observed in many Gamma-ray Bursts (GRBs) has been interpreted as either being fueled by fallback onto a newly formed black hole, or by the spin-down luminosity of an ultra-magnetized millisecond neutron star. If the latter model is assumed, GRB X-ray afterglow light curves can be analytically reproduced. We fit a sample of GRB X-ray plateaus, interestingly yielding a distribution in the magnetic field versus spin period (B-P) diagram consistent with $B\propto P^{7/6}$. This is expected from the well-established physics of the spin-up line minimum period for Galactic millisecond pulsars. {\bf From} the normalisation of the relation we obtain perfectly matches spin-up line predictions for the expected masses ($\sim 1 M_{\odot}$) and radii ($\sim 10 {\rm ~km}$) of newly born magnetars, and mass accretion rates consistent with GRB expectations of $10^{-4} M_{\odot}/{\rm s}  <\dot{M}< 10^{-1} M_{\odot}/{\rm s}$. Short GRBs with extended emission (SEE) appear towards the high period end of the distribution, while the long GRBs (LGRBs) towards the short period end. This result is consistent with spin-up limit expectations where the total accreted mass determines the position of the neutron star in the B-P diagram. The {\bf P-B} distribution for LGRBs and SEE are statistically different, further supporting the idea that the fundamental plane relation \citep{dainotti16c,Dainotti2017}, {\bf a tri-dimensional correlation between the X-ray Luminosity at the end of the plateau, the end time of the plateau and the $1-s$ peak luminosity in the prompt emission}, is a powerful discriminant among those populations. Our conclusions are robust against suppositions regarding the GRB collimation angle and magnetar breaking index, which {\bf shifts} the resulting magnetar properties parallel to the spin-up line, and strongly support a magnetar origin for GRBs presenting X-ray plateaus. 
\end{abstract}


\section{Introduction}
\label{sec:intro}
Gamma ray bursts (GRBs) have been historically classified as short (SGRBs) with $T_{90}<2 s$ ($T_{90}$ is the time
{\bf it takes the burst to emit} between $5\%$ and $95\%$ of its isotropic emission) and long (LGRBs) with $T_{90}>2s$. With the afterglow discovery \citep[e.g.][]{Costa1997}, this classification found physical grounds in the evidence of two distinct progenitors for the LGRBs, associated with the gravitational collapse of massive stars, and the SGRBs recently confirmed to be associated with compact binary coalescences \citep{Abbott2017}. However, the continuously increasing sample of discovered GRBs is suggesting the existence of a more complex classification scheme that may reflect further differences in the progenitor properties. An example of a possible sub-population is represented by those LGRBs that lack an associated supernova (SN), e.g. GRB 060505 \citep{fynbo06,dellavalle06,Gal-Yam2006,Fruchter2006} (Tutukov \& Fedorova 2007) 
 X-ray flashes (XRFs) are also an empirically-defined subclass of LGRBs,  characterized by a greater fluence in X-rays ($2-30$ keV) than in the $\gamma$-ray band ($30 - 400$ keV). However, whether they can be considered as full-fledged members of the LGRB population is still debated, e.g. \citep{Sakamoto2008,Stratta2006}. 

The most significant failure of the simple long-short dichotomy is represented by a subset of SGRBs, labelled short GRBs with extended emission (SEE, \citealt{norris2006,Norris2010}), that show a soft and temporally extended component immediately after the characteristic short/hard initial peak.  
SEEs have $T_{90}>2$ s like LGRBs, while their spectral properties are similar to those of SGRBs. 
The puzzling nature of SEE possibly encodes important pieces of information that are still missing in our understanding of GRB progenitors. 

Despite the diversity in GRB phenomenology, common features may be identified in their afterglow lightcurves. 
In particular, a large fraction of X-ray afterglows show a plateau emission in the early stages (i.e. less than a day since the burst, \citealt{Obrien06,Nousek2006,Zhang2006,sakamoto07}).
Plateaus are followed by a light curve steepening in line with the external-shock fireball predictions \citep[e.g.][]{Sari98,sari1999}.    
A general consensus exists on attributing the plateaus to prolonged energy injection by the central engine, although a conclusive explanation for all their properties is still lacking. Substantial debate concerns whether the energy injection is due to a newborn, spinning down neutron star (NS) or to the fallback of matter onto a newly formed black-hole (BH). Recently, \cite{Li2017} suggested the existence of a mixed population of central engines for a sample of GRBs with plateaus, with at most $\sim$ 20\% of the plateaus being consistent with the maximum spin energy of a NS, the rest requiring a more energetic central engine (i.e., a BH remnant). A comparison of our results is included in the discussion.

{\bf Independently} of the physical interpretation of plateaus, adopting an empirical approach \citealt{Willingale2007} (hereafter W07) found that Swift GRB prompt+afterglow light curves computed in the 0.3-10 keV energy range \citep{OBrien2006}, can be described with the same analytical expression. 
Adopting the W07 functions to identify the GRB plateau phase, \cite{Dainotti2008,dainotti2010,dainotti11a,Dainotti2013a,dainotti15} found an anti-correlation between rest-frame time at the end of the plateau, $T_{a}$, and its corresponding X-ray luminosity, $L_a$. The relevance of the so-called Dainotti relation is twofold: it provides important constraints about the physical properties of the central engine and enables the use of GRB afterglow correlations as cosmological tools \citep{cardone09,cardone10,postnikov14,Dainotti2013b}.
After correcting this relation for selection bias and redshift evolution \citet{Dainotti2013a} showed that the intrinsic slope of the correlation is $\approx -1$, which could be interpreted as a fixed energy reservoir during the plateau phase. This possibility was explored both in the context of the fallback mass surrounding the BH \citep[e.g.][]{Cannizzo2009,cannizzo2011} and for {\bf the spin-down} of a newly born magnetar  \citep{dallosso2011,Rowlinson2010,OBrien2012,Nemmen2012,Rowlinson2013,rowlinson14,rea15}. To better compare the Dainotti relation with the existing physical models it is necessary to disentangle properties of the different classes. Indeed, Dainotti et al. (2017c) showed that for the GRBs with a SNe associated (LONG-SNe) spectroscopically, the energy reservoir for the plateau is not constant. Here the LGRBs for which the SNe have not been seen (LONG-NO-SNe) are not included in the sample for this reason. 
More recently, Dainotti et al. (2016a) extended the 2D Dainotti relation into a new 3D relation by adding the peak luminosity in the prompt emission, $L_{peak}$. This relation, thus, identifies a fundamental plane in a three dimensional space parameter where now also the prompt emission properties are included. 
To isolate class-specific properties, the above studies defined observationally homogeneous samples of GRBs: XRFs, LONG-SNe, SEEs and LONG-NO-SNe-NO-XRFs. Indeed, the use of homogeneous samples usually helps to reduce the scatter of correlations \citep{dainotti2010,dainotti11a,delvecchio16}, a common practise also valid for prompt correlations \citep{Yonetoku2004,Amati2009,dainotti16c} and prompt-afterglow correlations \citep{Dainotti11b,Dainotti2015b}. 
In particular, Dainotti et al. (2016a) identified a "gold" sample of LGRBs with good data coverage and relatively flat plateaus, not associated with SNe nor classified as XRFs or SEE. This sample has the smallest intrinsic scatter and thus defines a “gold” fundamental plane. 
Later, Dainotti et al. (2017) computed the distance distribution from the gold fundamental plane for the different GRB categories. {\bf They} found that the SEEs are the only class statistically different from the gold sample, while all other LGRBs (i.e. XRFs, LONG-NO-SNe) are consistent with the gold plane. 

Using the same homogeneously defined GRB sample as in Dainotti et al. (2016c,2017a), {\bf we here show} that if we assume the magnetar spin-down model, optimal fits result for the X-ray plateaus. The physical parameters of these fits yield initial periods, P, and magnetic field strengths, B, for the magnetars in question at the start of the plateau phase. When plotted on our B-P plane, these parameters surprisingly fall exactly within the region predicted by the spin-up line of accreting NSs in Galactic binary systems for mass accretion rates $10^{-4} {\rm M}_\odot/{\rm s}~< \dot{M} <0.1$ M$_\odot$/s, as expected in GRBs. Thus, it appears that a minimum NS period determined by the magnetically coupled accretion of material from a progenitor, explains the P and B values required to accurately model the X-ray plateaus. We see a self-consistent physical interpretation for the X-ray plateaus arising from spinning down NS.

In section \S \ref{sec:model} we introduce the magnetar spindown model, in section \S \ref{sec:sample} we describe the sample of GRBs that we considered in this work and the data analysis is presented in \S \ref{sec:analysis}. 
Results are then showed in \S \ref{sec:results}, and discussed in \S \ref{sec:discussion}, with our conclusions being presented in \S 7.

\section{The magnetar x-Ray afterglow model}
\label{sec:model}
Within the fireball model the NS spins down due to {\bf the emission of} magnetic dipole radiation, releasing a luminosity $L_{\rm sd}= I \Omega \dot{\Omega}$, where $\Omega = 2 \pi \nu$ is the NS spin rate and $I$ is its moment of inertia.

Generally, the NS spindown implies the relation $\dot{\Omega} \propto \Omega^n$, where the `braking index' $n=3$ corresponds to ideal MHD and $n \leq 3$ to the presence of non-ideal effects
\footnote{\bf A braking index $n>3$ is possible in new-born NSs if the spin down is caused also by additional energy losses, e.g. gravitational wave emission.  We note that $n>3$  has also been suggested for old pulsars (REF?). }. 

In the former case, we adopted the spindown luminosity \citep{Spitkovsky2006} 

\begin{equation}
\label{eq:Spit06}
L_{\rm sd} = \frac{\mu^2}{c^3} ~\Omega^4 \left(1+ \sin^2\theta\right) \, ,
\end{equation}
 where $\mu = B R^3/2$ is the magnetic dipole moment, $B$ the B-field strength at the magnetic pole, $R$ the NS radius and $\theta$ the angle between the magnetic and rotation axes. Eq. \ref{eq:Spit06} can be generalized to a non-ideal case, e.g., by adopting the parametrization proposed by \citet{Contopoulos2006}

\begin{equation}
\label{eq:CS06}
L_{\rm sd}^{({\rm N-I})} = L_{\rm sd} \left(\frac{\Omega}{\Omega_i}\right)^{-2 \alpha} \, \end{equation}
where ``N-I" stands for non-ideal case and {\bf the shorthand} $Q_i$ stands for {\bf any generic} quantity $Q$ {\bf evaluated} at the initial time. Here, $0 < \alpha < 1$ and the braking index is $n =3 -2\alpha$. Eq. (\ref{eq:CS06}) is obtained by postulating that, due to resistive effects in the magnetosphere, an increasing fraction of the magnetic flux from the NS becomes open to infinity as the NS spins down (e.g. \citealt{Contopoulos2006}, and references therein for an in-depth discussion). For fixed $B$ and $\Omega$, this leads to a stronger spindown than in (\ref{eq:Spit06}). 
Solving Eq. (\ref{eq:CS06}) for $\Omega$, the spindown luminosity as a function of time is,

\begin{equation}
\label{eq1}
L_{\rm sd}(T)=\frac{L_{\rm sd,i}}{\left[1+\left(1-\alpha\right) \displaystyle \frac{T}{\tau_i}\right]^{\displaystyle \frac{2-\alpha}{1-\alpha}}}~ = ~\frac{E_{\rm spin,i}}{\tau_i \left[1+(1-\alpha) \displaystyle \frac{T}{\tau_i} \right]^{\displaystyle \frac{2-\alpha}{1-\alpha}}},
\end{equation}

where $\tau = \Omega/(2 \dot{\Omega})$ is the spindown timescale,  $E_{\rm spin} = (1/2) I \Omega^2$ is the NS spin energy.

Following \citealt{dallosso2011}, we consider the energy evolution in the relativistic external shock by writing the balance between radiative losses and energy injection from the spinning down magnetar, in the observer's frame

\begin{equation}
\label{eq:budget}
\frac{dE}{dT} = L_{sd}(T) - k^\prime \frac{E}{T}~
\end{equation}
where $T$, the observer's time, is related to $t$, the time in the engine frame, via $dT =(1-\beta)dt\approx dt/2 \Gamma^2(t)$, $\Gamma(t)$ being the shock bulk Lorentz factor. The coefficient $k^\prime = 4\epsilon_e($dln$t$/dln$T$) depends on the electron energy fraction ($\epsilon_e$) and the dynamical evolution of the shock (dln$t$/dln$T$). For simplicity, we {approximate} $k^\prime$ as a constant, although in principle both $\epsilon_e$ and the logarithmic derivative may change with time. {\bf In particular, in the case of an ISM or wind medium, dln$t$/dln$T$ has a maximum of $1/2$ or $1$, respectively, for a nearly adiabatic shock, and a minimum of $1/(4+k)$ or $1/(1+k)$, respectively, for a passively ``cooling" one}. Thus, it may gradually change with time, as conditions in the shock evolve. 
The second term in Eq. (\ref{eq:budget}) represents radiation losses, i.e. the shock luminosity $L(T)=k'E(T)/T$. A solution of Eq. (\ref{eq:budget}) thus determines the lightcurve of the external shock. Finally, we have considered two possibilities for the angular pattern of emission from the spinning down NS. In the ideal case, magnetic dipole radiation is almost isotropic, 3/2 more luminosity being emitted along the magnetic equator than on average, and zero luminosity along the magnetic axis. However, far from being ideal, the post-GRB environment in which the NS is born may favour beaming of the outgoing MHD wind, even within a solid angle comparable to that of the fireball (e.g. \citealt{Duffell2015}). In the isotropic case, the energy injection term in Eq. (\ref{eq:budget}) must be reduced by a factor $f_b=(1-cos \theta_{jet})$, where $\theta_{jet}$ is the fireball jet half-opening angle, while in the beamed case all of the NS luminosity is channelled along the jet and reaches the external shock.  
{\bf We note that this is the first time that a non-ideal modelling of spindown magnetar is fitted to the afterglow data. }

\section{The sample} \label{sec:sample}

Following Dainotti et al. (2008, 2010, 2011a, 2013a, 2015a), we used the W07 model to identify the plateau phase for 176 X-ray afterglows detected by {\it Swift} BAT and XRT from January 2005 until July 2014 with known redshifts (both spectroscopic and photometric). We considered only those GRBs with robust redshift estimation taken from \citealt{xiao09} and from the Greiner web page \footnote{\url{http://www.mpe.mpg.de/~jcg/grbgen.html}}. 
We took the light curves from the Swift web page repository \footnote{\url{http://www.swift.ac.uk/burstanalyser}} and we followed Evans et al. (2009) for the evaluation of the spectral parameters.

The W07 model is defined as : 
\begin{equation}
f(t) = \left \{
\begin{array}{ll}
\displaystyle{F_i \exp{\left ( \alpha_i \left( 1 - \frac{t}{T_i} \right) \right )} \exp{\left (
- \frac{\tau_i}{t} \right )}} & {\rm for} \ \ t < T_i \\
~ & ~ \\
\displaystyle{F_i \left ( \frac{t}{T_i} \right )^{-\alpha_i}
\exp{\left ( - \frac{\tau_i}{t} \right )}} & {\rm for} \ \ t \ge T_i \\
\end{array}
\right .
\label{eq: fc}
\end{equation}

The complete light curve $f_{tot}(t) = f_p(t) + f_a(t)$ contains two sets of four parameters $(T_{p},F_{p},\alpha_p,\tau_p; T_{a},F_{a},\alpha_a,\tau_a)$ that describe the prompt, ($f_p$), and the afterglow, ($f_a$), emission. 
The parameter $\alpha_{i}$ is the temporal power law decay index and the time $\tau_{i}$ is the initial rise timescale. In our analysis $\tau_a$ is free to vary. The transition from the exponential to the power law occurs at the point $(T_{i},F_{i}e^{-\tau_i/T_i})$.  
The time at which $f_p=f_a$ is the $T_{start}$ of the plateau. We excluded those GRBs for which the fitting procedure fails or when the determination of 1$\sigma$ confidence interval does not fulfill the Avni prescriptions \citep{Avni}. To select an observationally homogeneous class of sources, we adopt the Dainotti et al. (2016a) homogeneous samples of SEE and LGRBs without XRFs, SEE and GRBs associated with SNe. 
In Dainotti et al. (2016a), the selection for the SEE sample is taken from the catalogs of \citealt{norris2006,Levan2007,Rowlinson2013,Norris2010}. To select only the LONG-NO-SNe GRBs, all the GRB-SNe which follow the Hjorth \& Bloom (2011) classification were not considered. Similarly, all LGRBs consistent with the XRFs definition were removed. 
These criteria reduce the sample from 176 to $122$ long GRBs. 
Following Dainotti et (2016c), we used only the gold sample (40 GRBs), namely GRBs which have at least $5$ points at the beginning of the plateau and the angle of the plateau is less than $41$ degrees. {\bf Our sample of selected GRBs cover a huge redshift range ($0.033<z<9.4$). With these two selection criteria of the minimum number of points in the plateaus and the steepness of the plateau itself the sample of 122 LGRBs presenting plateaus is then reduced further to 40 GRBs. This sample presents clear observational features of the plateau emission that can lead to a better understanding of the emission process underlying it. The reason why we choose $41$ degrees depends on the fact that the distribution of the angles follow a Gaussian distribution with a $10\%$ outliers beyond $41$ degrees. For details about the distribution see Fig. 10 of Appendix B in Dainotti et al. 2017b.}

\section{Data analysis}\label{sec:analysis}

We compute the GRB bolometric luminosity $L(t)$ as follows:
\begin{equation}
L(t)= 4 \pi D_L^2(z) \, F_X (E_{min},E_{max},t) \cdot \textit{K} \cdot (1-cos\theta_{jet}),
\label{eq: lx}
\end{equation}
where $F_X (E_{min},E_{max},t)$ is the 0.3-10 keV X-ray flux at any time from the time start of the plateau until the end of the observed light curves and $K$ is the K correction in {\bf the $1-10,000$ keV band} \citep{Bloom2001}. 

We assume that the afterglow emission is collimated, hence the $(1-cos\theta_{jet})$ factor in Eq. (\ref{eq: lx}). Given the large uncertainties affecting $\theta_{jet}$ from the multi-wavelength afterglow data analysis, we adopt here the method presented by \citealt{Pescalli2015} {\bf in which $\theta_{jet}$ can be estimated by using the $E_{peak}-E_\gamma$ relation (Ghirlanda et al. 2004) and the $E_{peak}-E_{iso}$ (Amati et al. 2002), where $E_{peak}$ is the energy at which the prompt emission $EF(E)$ spectrum peaks and $E_\gamma$ and $E_{iso}$ are the burst radiated energy assuming a collimated or an isotropic geometry, respectively.}

Dainotti et al. (2013a), following the approach of Efron \& Petrosian 1999 \citep{Efron1992}, have showed that the
plateau luminosity and duration evolve with redshift as $(1+z)^{k_i}$ where $k_L=-0.05$ and $k_T=-0.85$
for the luminosity and duration, respectively. We thus applied these corrections to the
bolometric luminosity and rest-frame time and used these new lightcurves 
{\bf in the fit with the} analytical expression of the magnetar model. 
More specifically, we created new de-evolved variables
indicated with the symbol ${'}$ as local (as they would appear at $z=0$) luminosities and
times. $L_i^{'}=L_i/((1+z)^{k_L})$ and $T^*_i/((1+z)^{k_T})$ are constructed for each data
point $i$ of the lightcurves starting from the $T_{start}$ of the plateau. 
{\bf Previous analyses have been performed with the evolutionary corrections \citep{rowlinson14,rea15} by employing MonteCarlo simulations in the parameter spaces of magnetic field, $B$, and spin periods, $P$, allowed by the magnetar model and the observed Luminosities and the time duration of the plateau emission. These previous approches have not been performed by fitting the lightcurves. The sample adopted here is different from the samples adopted in previous studies since it relies on the gold sample and on the SEE sample.} 
To the best of our knowledge, this is the first time in the literature that this evolutionary
correction has been applied directly
{\bf to the rest-frame lightcurves in a fitting procedure.}

This ensures that the set of parameters derived, (E, k, P and B), are no longer affected by selection
bias or redshift evolution, and thus, they are intrinsic to the physics of GRBs.
If for example, these sets of the parameters followed a particular evolutionary path and/or were subject to
observational selection biases, we would obtain a set of parameters also affected by such biases and
hence leading to inferences not necessarily inherent to the systems being treated. In that case, the
distribution of these inferred variables would represent physical conditions associated to the relevant
biases rather than to the intrinsic physics being sought. The rescaling applied prevents any such case. 

We test the luminosity lightcurves against the magnetar model exploring both the isotropic and beamed emission. In the isotropic case, the magnetar energy injection is reduced by a factor $f_b =(1-cos\theta_{jet})$, while in the beamed case all of the NS luminosity is channelled along the jet and reaches the external shock. For each scenario, we test the case of an ideal magnetic dipole with braking index $n=3$, i.e. $\alpha=0$ (see \S \ref{sec:model}) and the case with $n=1.2$  ($\alpha=0.9$). 
In the fit we treat the energy at the beginning of the plateau, $E(T_{start})$, as well as $B$, $\Omega_i$ and $k'$ as free parameters which are varied to optimize the fit.  
We considered statistically acceptable fits those with p-value$>$0.05. (i.e. for which there is $\leq$ 5\% chance of obtaining a set of measurements at least this discrepant from the model, assuming the model is true). 
In GRB 0812221, GRB140506A and 060614A we removed small flares or bumps observed at late times (i.e. $>1-10$ days, that is well after the plateau's end time), {\bf not to} affect the best parameter estimates with an unrelated
phenomenology. For GRB 061222A, GRB 070508 we accepted lower p-value thresholds down to 0.005, since the overall temporal behavior is well represented by the magnetar model over several order of magnitudes in time and the low p-value may be due to small flux variations due to a persistent flaring activity.
{\bf Results are quoted in Table 1 where the number of GRBs with acceptable fit is labelled with $N_{good}$.
For a different approach in which the data determine the breaking index see Lasky et al. 2017. In this work Lasky et al. 2017 found that one GRB has $n = 2.9$, and the other $n = 2.6$, both with uncertainties of 0.1 at the one-sigma confidence level.}  

\begin{sidewaystable}
\centering
\caption{LGRBs and SEEs {\bf fits} with the magnetar models. {\bf The third column quotes the number of GRBs that could be fit by the corresponding model with spin period $P>0.51$ ms ($N_{good}$), as explained in \S \ref{sec:discussion}, over the total number of GRBs analyzed $(N_{tot})$. } The mean values of magnetic field $B$, spin period $P$ and coefficient $k'$ (Eq.\ref{eq:budget}) are quoted for two different braking indexes $n=3-2\alpha$ and different energy release geometry (collimated and isotropic) of the magnetar. The Pearson correlation coefficients $r$ in the $B-P$ log-log plane and p-value for the null hypothesis are quoted for LGRBs, SEE and LGRBs+SEEs. 
\label{tab:general}}
\begin{tabular}{ l | l |  c | c | c | c | c | c || c | c | c |  c | c | c }
\hline
 & & \multicolumn{6}{c}{NS iso}  & \multicolumn{6}{c}{NS collimated}\\
\hline
sample & model & $N_{good}/N_{tot}$ & $<B>$ & $<P>$ & $<k'>$ & $r$ & p-val &  $N_{good}/N_{tot}$ & $<B>$ 	& $<P>$ & $<k'>$ & $r$ & p-val \\
	   & 		& 				  & $10^{14}$G & ms & 		& 	 & 	   & 		& $10^{14}$G & ms & 		& 	 &  	\\
\hline
LGRB & $\alpha=0$ 	& 19/40 & $6\pm4$ 		& $1.1\pm0.9$ & $0.5\pm0.4$ & 0.4 & 0.09 & 33/40 & $64\pm45$ & $9\pm6$ & $0.5\pm0.3$ & 0.42 & 0.01\\		
	 & $\alpha=0.9$ & 26/40	& $5.4\pm4.7$ 	& $1.0\pm0.8$ & $0.5\pm0.4$ & 0.49 & 0.01 & 37/40 & $53\pm40$ & $9\pm6$ & $0.5\pm0.3$ & 0.62 & 4e-5 \\
\hline
SEE	& $\alpha=0$ & 10/11  & $33\pm20$ & $4.0\pm2.5$ & $1.3\pm1.2$ & 0.48 & 0.16 & 10/11 & $258\pm148$ & $27\pm14$ & $1.1\pm1.1$ & 0.32 & 0.4\\
	& $\alpha=0.9$ & 10/11& $27\pm13$ & $5.0\pm4.1$ & $0.7\pm0.5$ & 0.22 & 0.53 & 10/11 & $209\pm111$ & $34\pm28$ & $0.5\pm0.4$ & 0.17 & 0.63\\
\hline
\hline
LGRB+SEE	& $\alpha=0$ & 29/51  & -- & -- & -- & 0.75 & 3e-6 & 43/51 & -- & -- & -- & 0.65 & 2e-6\\
	& $\alpha=0.9$ & 36/51& -- & -- & -- & 0.74 & 2e-7 & 47/51 & -- & -- & -- & 0.71 & 1e-8\\
\hline

\end{tabular}
\end{sidewaystable}

\section{Results} \label{sec:results}

\begin{figure}[ht]
\centering
\includegraphics[scale=0.35]{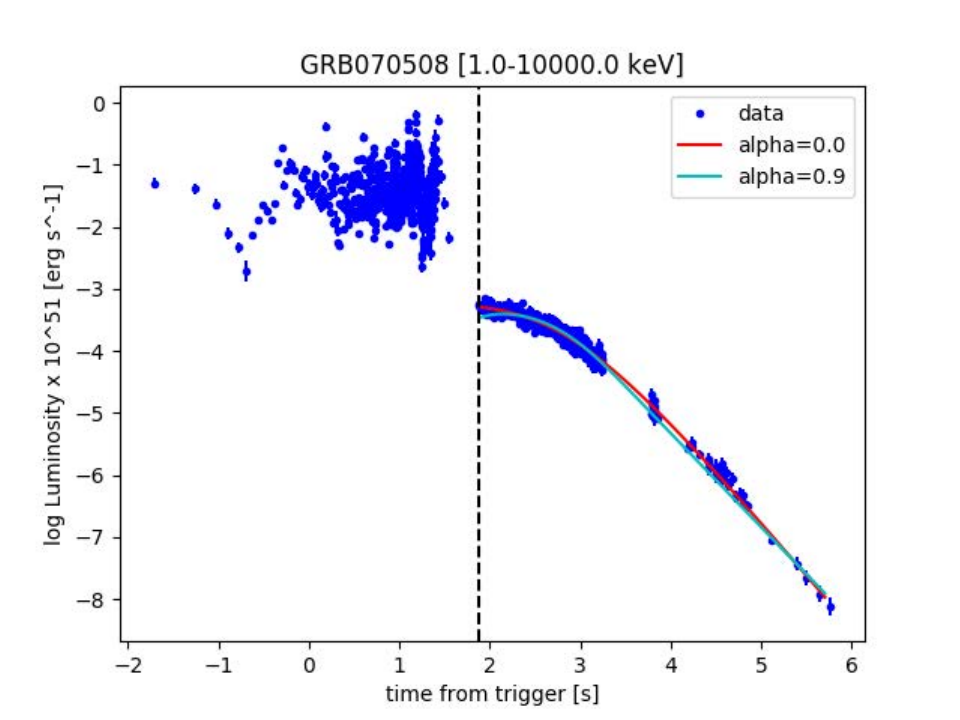}
\includegraphics[scale=0.35]{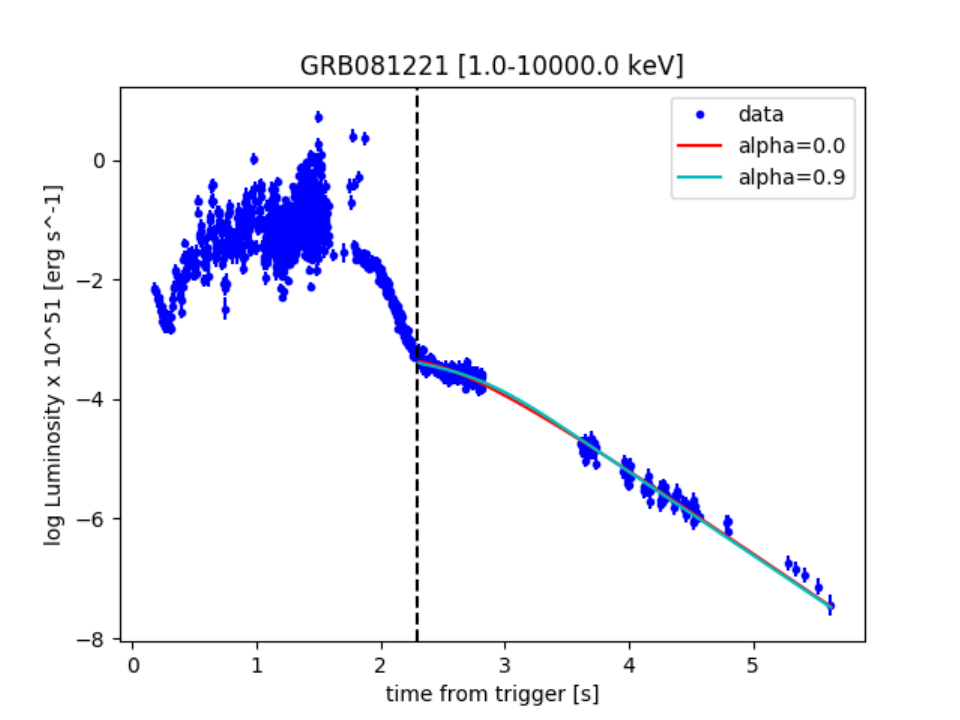}
\includegraphics[scale=0.35]{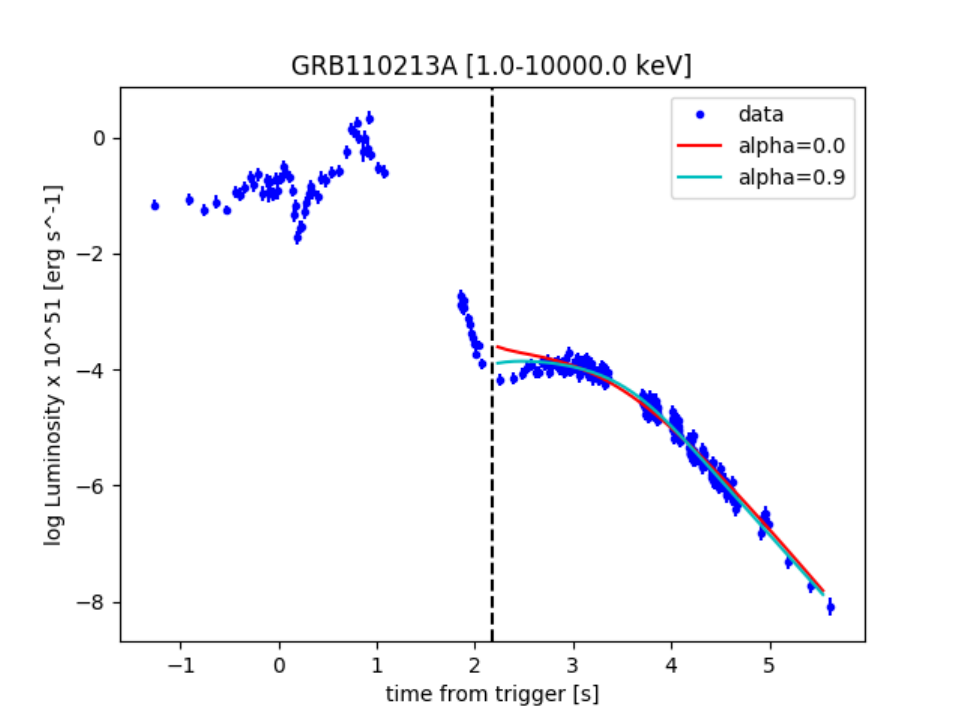}
\includegraphics[scale=0.35]{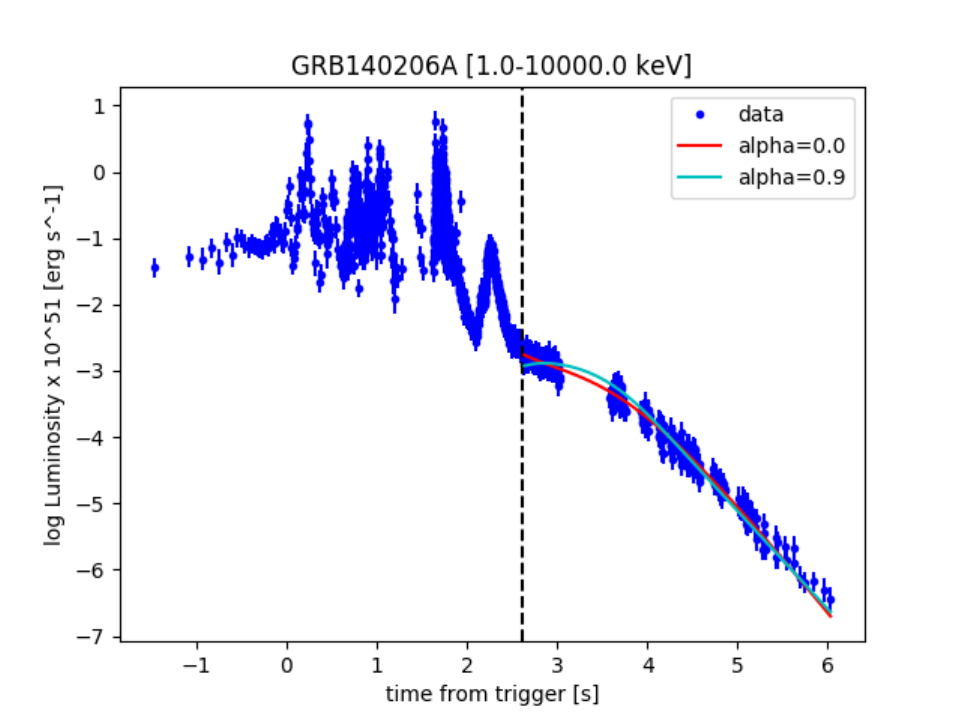}
\includegraphics[scale=0.35]{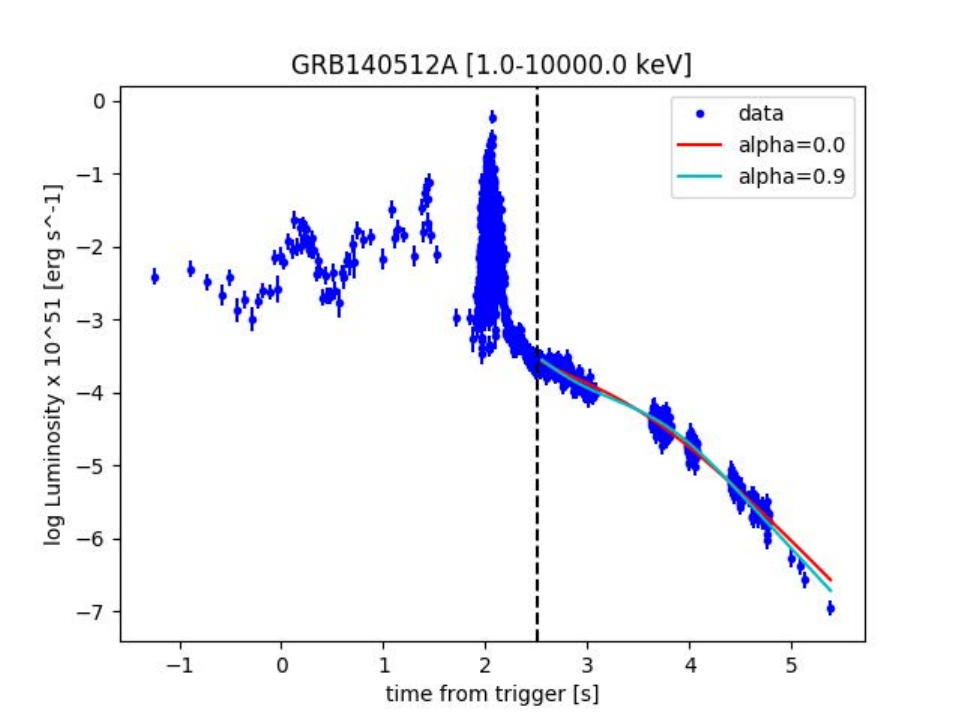}
\includegraphics[scale=0.35]{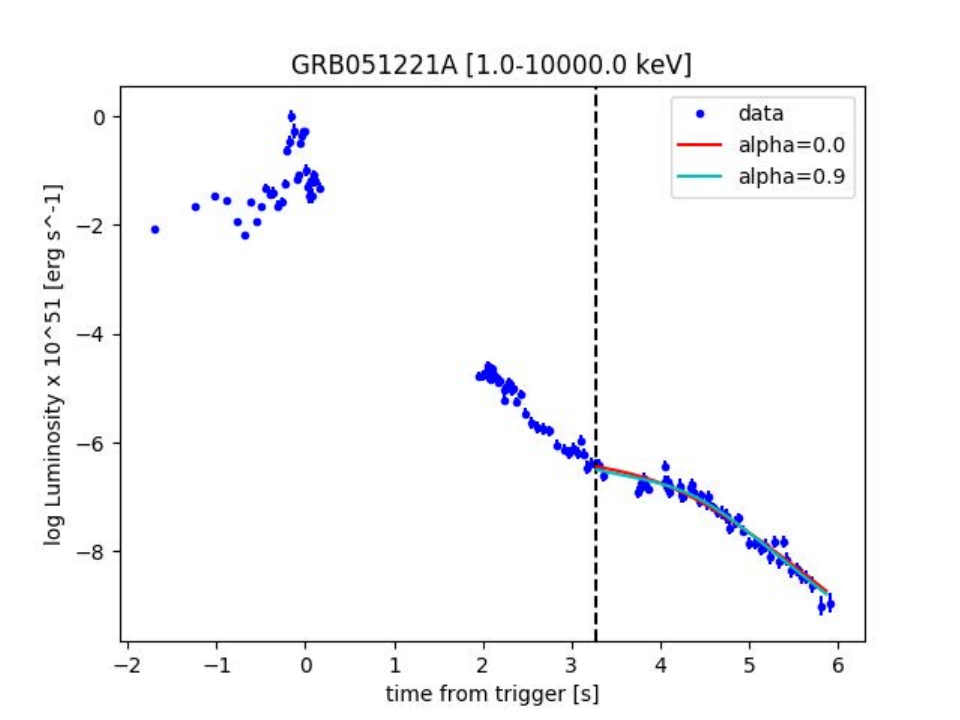}
\includegraphics[scale=0.35]{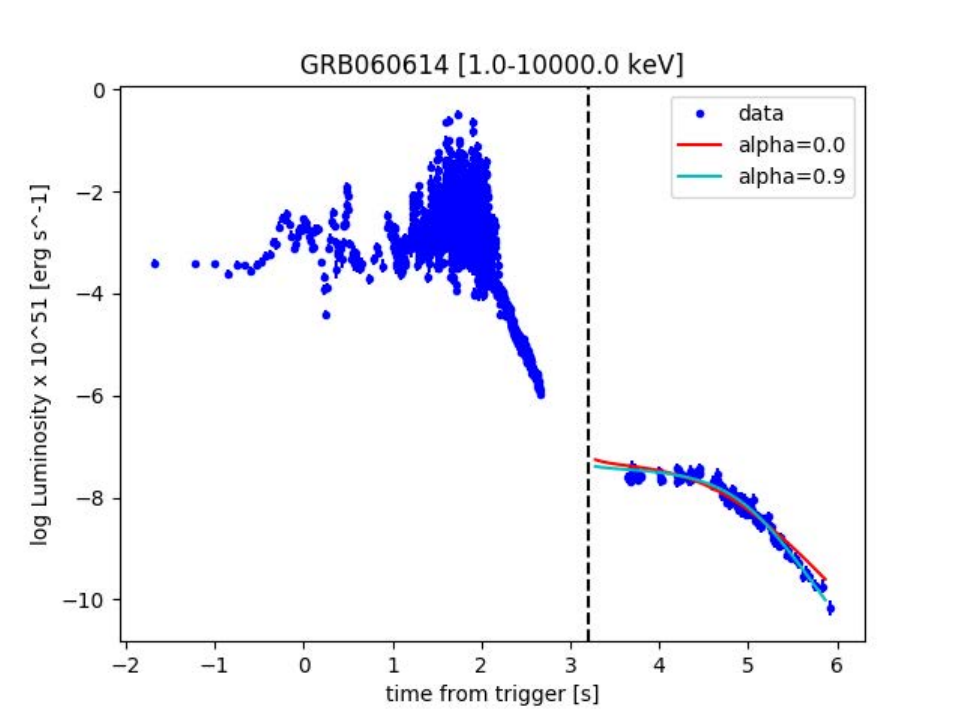}
\includegraphics[scale=0.35]{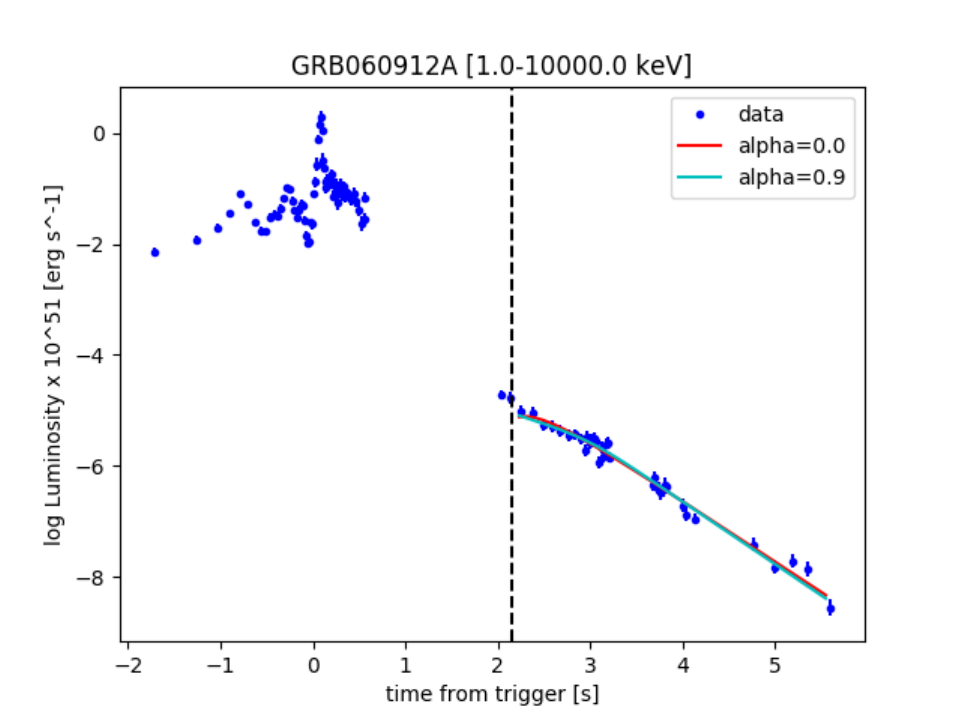}
\caption{Examples of five LGRBs and three SEEs Swift/XRT X-ray light curves (blue dots) for which we have high statistics. 
Red lines show the pure dipole radiation model with $\alpha=0$, while cyan lines are the ones with $\alpha=0.9$. In these plots the magnetar energy release was assumed to be collimated (see Discussion). \label{fig:good}}
\end{figure}

The toy model described in \S \ref{sec:model} provides excellent fits to the X-ray lightcurves for almost all the GRBs in our sample. Figure \ref{fig:good} shows eight examples {\bf (five long GRBs and three short with extended
emission ones)} taken from among those with the largest available statistics in the light curve sampling, where it can be seen that the best fit parameters for the proposed model indeed allow for an accurate representation of the empirical light curves. Independently on the assumed geometry in the magnetar energy release, we find that LGRBs 060607A, 081028A and 100906A cannot be fitted by any assumptions on $\alpha$. 

The fraction of well fitted systems further increases if we introduce a modification to the pure radiating dipole, i.e. by allowing the braking index to be $n<3$. 
The number of GRBs with a good fit, and the average values of $B$, $k'$ (see Eq.\ref{eq:budget}) and $P$ in the different scenarios are summarized in Table \ref{tab:general} {\bf where the number of GRBs that could be fit by the corresponding model with spin period $P>0.51$ ms ($N_{good}$), as explained in \S \ref{sec:discussion}, is quoted over the total number of GRBs analyzed $(N_{tot})$. }  
In the case of an isotropic NS emission a significant fraction of the {\bf LGRBs analyzed show} unphysical values of $P$, below a minimum of 0.51 ms discussed in Sec. 6. 
Larger spin periods (all above 0.51 ms) are recovered by assuming a beamed scenario. 
By assuming an isotropic magnetar the best agreement with the full sample of LGRBs is obtained with $\alpha=0.9$ for which statistically acceptable fit were found for 37/40 LGRB. Assuming an ideal magnetic dipole radiation ($\alpha=0$), statistically acceptable fits are obtained for 33/40 LGRB. For the beamed magnetar scenario, we find that the best agreement with LGRBs is obtained again by assuming $\alpha=0.9$ for which we find statistically acceptable fit for 37/40 LGRB (Table \ref{tab:a09coll}). For an intermediate value of $\alpha=0.3$ we found a good agreement for 36/40 LGRBs. For a more realistic value of $\alpha=0.3$, statistically acceptable fit were found for 36/40 LGRB. For the SEE, statistically acceptable fits are obtained assuming any $\alpha$ and geometry with the exception of GRB 060614 that could be fitted only by assuming $\alpha=0.9$. 
Contrary to LGRBs, the initial $P$ for SEEs is always found above the minimum value, both for a collimated and isotropic energy release from the magnetar. 

\begin{figure}
\gridline{\fig{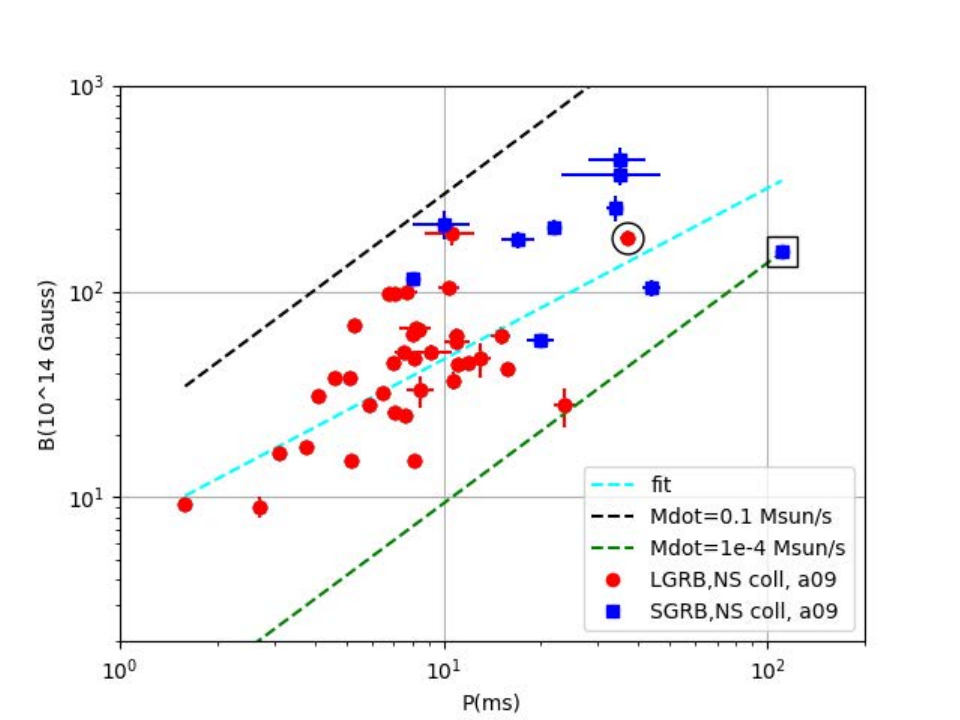}
{0.75\textwidth}{}
        }
\caption{The magnetic field vs spin period assuming the spindown luminosity of the magnetar is entirely beamed within the jet opening angle (data points) for a braking index n=2.1 ($\alpha=0.9$). The long GRBs are marked as red circles while the SEEs are blue squares. 
Dashed lines indicate the expected $B-P$  relations from accreting NS for an accretion rate of $0.1 M_{\odot}/s$ (black) and $10^{-4} M_{\odot}/s$ (green) and the best-fit relation (cyan).  The two framed datapoints indicates the  long GRB 070208 (circle) and the peculiar 
GRB 060614A (square). 
 \label{fig:pyramid}}
\end{figure}

The best fit values of $B$ and $P$ of LGRBs plus SEE in the log-log B-P plane show evidence of a correlation (Fig.\ref{fig:pyramid}). 
We compute the Pearson correlation coefficients and the corresponding p-value for the null-hypothesis for the sample of LGRBs, SEE and LGRBs+SEEs. Results are quoted in Table \ref{tab:general}.
By fitting our B vs P distribution for the case of a beamed magnetar and $\alpha=0.9$ (Fig.\ref{fig:pyramid}) with a power law we find: 
\begin{equation}
\log B=0.84+ \left(0.83^{+0.16}_{-0.18}\right) \log P
\end{equation}
where the 1$\sigma$ error on the slope is computed by taking into account errors on both variables \citep{Dagostini2005}. 

To test whether SEEs and LGRBs represent distinct populations, as suggested by Dainotti et al. (2017), we apply a Kolmogorov-Smirnov test (KS) and an Anderson-Darling test on the obtained $B$ and $P$ distributions for those GRBs with acceptable fit and with $P>0.51$ ms (see next section).  
Results are quoted in Table \ref{tab:stat} and provide robust evidence of two distinct populations. 

The LGRB 070208 is {\bf an} "outlier" that populates the SEE region. The light curve of this burst shows the presence of two peaks, with the first one much shorter than the second one. This could possibly indicate an SEE origin rather than a LGRB. {\bf However}, the widths of the two peaks and the fact that the second peak is spectrally harder than the first led to the conclusion that this burst is indeed a long one \citep{Markwardt2007}. Our results {\bf suggest} a possible unsolved ambiguity in the classification of this GRB and may indicate that indeed an SEE nature is more appropriate. 

{\bf A final GRB worth} mentioning is GRB 060614, which has both properties of a LGRB and a SGRB. Indeed, it is placed among the SEE population in the B-P plane, {\bf further} confirming its links to SGRB progenitors.

\begin{table}
\centering
\caption{Results from the KS and Anderson-Darling tests applied on the distribution of $B$ and $P$ for the LGRB+SEE sample. {\bf The third column quotes the number of GRBs that could be fit by the corresponding model with spin period $P>0.51$ ms ($N_{good}$), as explained in \S \ref{sec:discussion}, over the total number of GRBs analyzed $(N_{tot})$. } The KS statistics with the associated chance probability and the Anderson-Darling statistics with the associated significance level (an approximate significance level at which the null hypothesis for the provided samples can be rejected) are quoted for different assumptions on the magnetar model. \label{tab:stat}}
\begin{tabular}{ l | c | c |  c | c | c | c || c | c | c | c | c }
\hline
	&	&  \multicolumn{5}{c}{NS iso}  & \multicolumn{5}{c}{NS collimated}\\
\hline
 Distribution & $\alpha$ & $N_{good}/N_{tot}$	& KS & p-val & And.-Darl. & sign. &  $N_{good}/N_{tot}$	& KS & p-val & And.-Darl. & sign.\\
\hline
B	& 0.0 & 29/51 & 0.84 & $5\times10^{-5}$ & 11.2 & $7\times10^{-5}$ & 43/51 & 0.78 & $6\times10^{-5}$ & 11.3& $7\times10^{-5}$\\
P	& 0.0 & 29/51 & 0.74 & $6\times10^{-4}$ & 8.9 & $2\times10^{-4}$  & 43/51 & 0.71 & $3\times10^{-4}$ & 10.3& $1\times10^{-4}$ \\
\hline
B 	&0.9 &	36/51 &	0.92 & $2\times10^{-6}$	& 13.7 & $3\times10^{-5}$ &  47/51 & 0.82 & $1\times10^{-5}$ & 13.5& $3\times10^{-5}$ \\
P 	&0.9 &	36/51 &	0.81 & $5\times10^{-5}$	& 11.8 & $6\times10^{-5}$ & 47/51 & 0.74 & $1\times10^{-4}$ & 10.1& $1\times10^{-4}$ \\
\hline
\end{tabular}
\end{table}

\section{Discussion}\label{sec:discussion}

\subsection{Validity of the magnetar scenario}
We tested whether the magnetar model was consistent with all GRBs in our sample, and obtained accurate fits to the X-ray lightcurves for ($90\%$) of our selected plateaus.
Besides the successful fits, we checked the consistency of the resulting parameters with physical constraints on NS properties, in particular its maximum spin energy. 
The latter is set by the equation of state (EoS) of matter at supranuclear density, and depends on the NS mass and radius. It is $\lesssim 3 \times 10^{52}$ erg for a ``standard" NS, with a mass $\sim$ 1.4 M$_\odot$ and a 12 km radius, for which the minimum spin period is $\approx 1$ ms (Lattimer \& Prakash 2016, Bucciantini et al. 2007, 2009; Duffell et al. 2015). However, in more massive ($\gtrsim 2.1$ M$_\odot$) and compact (R $\approx 10$ km) NSs, the maximum spin energy can reach up to $\sim 10^{53}$ erg (e.g., Dall'Osso, Stella \& Palomba 2018 and references therein). 
{\bf Since in our fits we have assumed a fiducial NS with a mass of 1.4 M$_\odot$ and 12 km radius, a spin energy of 10$^{53}$ erg would correspond to} an {\it effective} minimum spin of $ 0.51$ ms.  
{\bf Thus, all spin periods $>0.51$ ms obtained in our fits are considered to be consistent with a NS central engine having a different mass and/or radius than the fiducial values.}

An additional source of uncertainty in the energy budget of the magnetar is the degree of anisotropy of its spindown luminosity (\S \ref{sec:model}). 

The spindown energy of a millisecond NS is carried out by a Poynting-flux/electron-positron wind, which is expected to be launched in a near isotropic fashion \citep[e.g.][]{usov92, Metzger2014, Fan2006}. 
However, if the magnetar is born at the center of a collapsar, the progenitor star could cause a collimation of the emitted energy, both during the prompt  \citep{Bucciantini2007,Bucciantini2009,Duffell2015} and the afterglow phase \citep{lu2014}. \citealt{Mazzali2014} argue for the magnetar wind to be close to isotropic in collapsars based on the energetics of the associated supernovae and on the assumption 
that the maximum spin energy of a NS is $\sim 2 \times 10^{52}$ erg. However, as pointed out above, the whole range of possible spin energy values leaves room for non-negligible beaming even in this case. {\bf Indeed, for a deeper understanding of this problem the formation of the Proto-Magnetar (PM) together with the
relevant physics and microphysics (Obergaulinger \& Aloy 2017) should be taken into account. Obergaulinger \& Aloy 2017 strongly suggest that (1) the
formation of a PM is preceded by the ejection of a highly-collimated
outflow, (2) the environment surrounding the PM is highly anisotropic,
with a large variance in properties from the direction of the rotational
axis to the equatorial region, and (3) the maximum rotational energy
that one may store (and, hence, constitute the energy reservoir for the
ulterior magnetar spin-down) can be significantly larger than the limit
suggested previously (Mazzalli et al. 2014). Both the collimated ejecta
and the anisotropy of the medium surrounding the PM may certainly have
an impact on the observed distribution of the spin-down energy. This theoretically explains why the preferred distribution obtained by our fit favour collimated energy.}

Table \ref{tab:general} shows that, for the isotropic and ideal case, 18 over 40 LGRBs would require an unphysical NS spin period, indicating a clear preference for a non-isotropic emitting magnetar, and a non-ideal spindown model ({\it i.e}, $n <3$). By assuming that the magnetar collimation is the same as the afterglow one, for the case with $\alpha=0.9$ only 4 out of 51 systems are not well fitted, and all of them have $P>0.51$ ms. We note that this result continues to be valid even for a milder degree of collimation, and for intermediate values of $\alpha$.

\subsection{The B vs P correlation and the $L_{peak}$ vs $P$ anticorrelation}

The most novel finding of our work is that, independent of the assumed particular magnetar parameters (i.e. collimation and braking index), there is evidence for a statically significant correlation in the $B$ vs. $P$ plane for all GRBs in our sample. 
A $B-P$ correlation is expected from the well established physics of the spin-up line for accreting NS in Galactic binary systems.
For a given $\dot{M}$, the NS reaches an equilibrium period $P_{eq} \propto B^{6/7}$ \citep[e.g.][]{Bhattacharya1991,Pan2013}.   

\begin{equation}
\label{eq:bp}
\frac{B}{10^{14} G} \approx 15 ~\left(\frac{P_{\rm eq}}{1 {\rm ms}}\right)^{7/6} 
\left(\frac{M}{1.4M_{\odot}}\right)^{5/6} \left(\frac{\dot{M}}{0.01  M_{\odot}{\rm /s}}\right)^{1/2} \left(\frac{R}{12~ {\rm km}}\right)^{-3} \, .
\end{equation}

For the fixed values of $M$ and $R$ used in our fits, we find that all our data in the $B-P$ plane are encompassed by two lines corresponding to the range of mass accretion rates $10^{-4}~ {\rm M}_\odot$/s $ <\dot{M} < 0.1$ M$_\odot$/s. 
In other words, the normalization of the $B-P$ relation obtained here perfectly matches spin-up line predictions for the magnetar model with mass accretion rates expected in the GRB prompt phase.
The latter are $\sim 11-14$ orders of magnitude higher than those inferred for accreting NSs where this physics was originally derived. 
Thus, it appears that the same basic physics is at play, but in powering a GRB, the accretion rates grow substantially as they are not the result of a companion shedding a fraction of its mass on a secular timescale, but rather, of a progenitor being swallowed whole by the nascent magnetar. 

{\bf Thus, the presence of a weak correlation in the B-P plane of our x-ray light curve fitted parameters with a slope
  consistent with that of eq.(8), serves as a clue that lead us to explore the spin-up line physics of Galactic
  milli-second pulsars in the context of GRBs. The correlation is necesarilly noisy due to the presence of a
  variety on opening angles and a range of radii, masses and detailed magnetic physics across the sample. The strong
  result is the bounding of the B-P parameters found by the two expected limits for $\dot{M}$ in eq.(8) in the
  context of GRBs.}

Interestingly, SEEs systematically populate the high period end of the distribution, while the LGRBs appear to have systematically faster spins (and lower B-fields).
This evidence can be interpreted in {\bf a} consistent way with the spin-up limit expectations, if SEEs were associated to progenitors yielding, on average, a smaller amount of accretion material than LGRB progenitors, e.g. binary NS mergers as opposed to collapsars. 
Indeed, the total accreted mass ($\Delta M$) determines the position of a millisecond pulsar in the B-P diagram \citep{Pan2013}, a greater $\Delta M$ corresponding to a larger transfer of angular momentum and, thus, a shorter spin period of the nascent magnetar. 

We have shown that the X-ray afterglow light curves can be fully described using a set of four parameters (k, E, B, P). The latter two, which are related to the neutron star properties are also correlated, as shown in Fig. 2. This
provides a physical foundation for the Dainotti 2D relation between $L_a$ and $T_a$. This
correlation carries its own intrinsic scatter, as expected from the spread in the B-P plane, which
 is only partially due to an intrinsic spread in the mass and radii of the neutron star,
but mostly due to a spread in the mass accretion rates. Thus a clear link
between the X-ray afterglow and the prompt energetics is established, as $\dot{M}$ is the key physical
parameter in determining the luminosity during the prompt phase. This last in turn explains
the 3D Dainotti relation where the X-ray afterglow parameters are linked to the prompt
$L_{peak}$; at fixed opening and viewing angle, and fixed mass and radius of the neutron,
$\dot{M}$ determines $L_{peak}$ and, together with the initial magnetic field, fixes the
position of the nascent magnetar along the particular spin-up line, which then
determines the resulting X-ray afterglow light curve.

If $\dot{M}$ can be considered as a proxy for the luminosity in the prompt phase (L$_{\rm peak}$), our result provides a natural connection between the properties of the plateaus and those of the prompt emission in GRBs. Observationally, one such correlation has been identified by Dainotti et al. (2017). 
Additionally, we would expect an anti-correlation between $P$ and L$_{\rm peak}$.
We found a weak indication of such an anti-correlation (the Pearson correlation coefficient is only $-0.33$), with a trend given by  L$_{\rm peak}\propto  P^{-0.27}$. However, due to the paucity of the data a definite conclusion cannot be drawn. 
Further work on this is currently under way and will be the subject of a subsequent publication.

\section{Conclusions}

In this work we analyze the plateau phase observed in the X-ray afterglows of two homogeneous samples of GRBs (LGRBs and SEEs) in the context of the magnetar paradigm, taking into account two angular patterns {\bf for} the magnetar energy release (i.e. collimated vs isotropic) and different 
values of the braking index $n$. The afterglow light curves can be accurately reproduced for the vast majority of our sample. 
The distribution of the best fit values of the magnetic field strengths $B$ and the spin periods $P$ for the SEEs and the LGRBs samples, are consistent with being originated from two distinct populations.
The evidence for SEE as constituting a separate population was first obtained by \cite{Dainotti2017} on more empirical grounds. The confirmation of the existence of two distinct GRB populations also on physical basis, has fundamental implications on the use of the GRBs correlations as cosmological tools, in the efforts toward a general standardization of these sources (see Dainotti et al. 2016, 2017).

It is extremely interesting that the magnetar parameters we find, when plotted in a B-P diagram, are neatly encompassed by the well-known spin-up line physics of $B \propto P^{7/6}$ of Galactic millisecond pulsars. This, for typical NS masses $\approx$ 1.4 $M_\odot$ and radii  $\approx$ 12 km, requires mass accretion rates $\approx 10^{-4} M_{\odot}/{\rm s} \leq \dot{M} \leq 0.1 M_{\odot}/{\rm s}$, typically expected in GRBs, which are about 11-14 orders of magnitude larger than typical in accreting pulsars. 
Thus, it appears that the same basic physics is at play even in the huge mass accretion rate that powers GRB.

We see LGRBs systematically requiring shorter periods, as expected from a larger total accreted mass resulting in more transfer of angular momentum than for the SEE GRBs.
This is consistent with interpreting the period of the magnetar as being fixed by the spin-up line, and also with our independent discovery of SEEs and LGRBs being distinct populations.
The B-P correlation is robust since it is independent {\bf of} specific assumptions made on details on the magnetar spin down. The collimation factor and the braking index mostly shift the resulting B vs P distribution parallel to the spin-up line. Our study therefore strongly supports the magnetar central engine origin for the X-ray plateaus.

\section*{Acknowledgments}
M.G.D. acknowledges the Marie Curie Program, because the research leading to these results has received funding from the European Union Seventh Framework Program (FP7-2007/2013) under grant agreement no. 626267. S.D. acknowledges support from NSF award AST-1616157. X.H. acknowledges financial assistance from UNAM DGAPA grant IN104517 and CONACyT. G.S. acknowledges EGO support through a VESF fellowship (EGO-DIR-133-2015).

\vspace{5mm}
\facilities{Swift(XRT and BAT)}

\begin{table}
\caption{Best fit results for the long (top) and short (bottom) GRB sample obtained by assuming a magnetar model with braking index $n<3$ with $\alpha=0.9$, and a collimated magnetar. \label{tab:a09coll}}
\begin{tabular}{| l | c | c | c| c | c | c | c | c |}
\hline
GRB 	& $t_0$ 	& $E(t_0)$	        & $\alpha$ & $k'$ & B 			& P 	& $\chi^2$ (d.o.f) & pval \\
	 	& s 	      & $10^{51}$ erg 		&         	& 	 & $10^{14}$ Gauss & ms &    			&  \\
\hline
060605 	& 157.7 	& 0.04	& 0.9 & $1.2 \pm 0.1 $		& $32.4 \pm 1.3$ 	& $6.5 \pm 0.1$ & 48.9(74) 	& 0.98933\\
060607A & ... & ... & ... & ... & ... & ... & ... & ... \\
060714 	& 375.5 	& 0.05 	& 0.9 & $0.27 \pm 0.04 $	& $50.5 \pm 3.5$ 	& $7.5 \pm 0.4$ & 33.7(44) & 0.86988\\
060814 	& 660.9		& 0.06	& 0.9 & $0.29 \pm 0.03 $	&$ 44 \pm 2$ 		& $11.1 \pm 0.4$ & 190.4(169) & 0.12353\\
060906 	& 262.6 	& 0.009	& 0.9 & $0.4 \pm 0.1 $		& $61 \pm 6$ 		& $15 \pm 1$ & 32.2(35) & 0.59990\\
060908 	& 123.3 	& 0.12	& 0.9 & $0.36 \pm 0.05 $	& $191 \pm 25$ 		& $10.6 \pm 1.9$ & 11.3(19) & 0.91099\\
061121 	& 226.6 	& 0.05	& 0.9 & $0.36 \pm 0.01 $	& $47.5 \pm 1.3$ 	& $8.1 \pm 0.1$ & 262.4(254) & 0.34388\\
061222A & 193.4 	& 0.2	& 0.9 & $0.40 \pm 0.02 $	& $17.7 \pm 0.5$ 	& $3.76 \pm 0.05$ & 399.3(335) & 0.00900\\
070208 	& 169.6		& 0.001	& 0.9 & $0.5 \pm 0.1$ 		& $182 \pm16$ 		& $37 \pm 2$ & 23.9(31) & 0.81215\\
070306 	& 457.5 	& 0.005	& 0.9 & $1.2 \pm 0.1 $	& $15.0 \pm 0.5$ 	& $8.1 \pm 0.1$ & 164.1(142) & 0.09815\\
070318 	& 6171.9	& 0.04	& 0.9 & $0.18 \pm 0.02 $	& $28 \pm 6$ 		& $23.7 \pm 1.8$ & 93.4(62) & 0.00603\\
070508 	& 76.0 		& 0.1	& 0.9 & $0.43 \pm 0.01 $	& $97 \pm 2$ 		& $7.1 \pm 0.1$ & 542.5(483) & 0.03121\\
070521 	& 98.7 		& 0.01	& 0.9 & $0.6 \pm 0.1 $	& $65 \pm 3$ 		& $8.4 \pm 0.2$ & 65.7(80) & 0.87562\\
070529 	& 165.3 	& 0.09	& 0.9 & $0.29 \pm 0.04 $	& $99 \pm 7$ 		& $7.7 \pm 0.6$ & 12.65811  (31) & 0.99856\\
080310 	& 999.3 	& 0.02	& 0.9 & $0.6 \pm 0.1 $	& $45 \pm 2$ 		& $11.9 \pm 0.4$ & 59.30765  (67) & 0.73683\\
080430 	& 374.0 	& 0.04 	& 0.9 & $0.10 \pm 0.01$	& $42 \pm 2$ 		& $15.7 \pm 0.6$ & 119.9(142) & 0.909952\\
081028A & ... & ... & ... & ... & ... & ... & ... & ... \\
081029 	& 377.7 	& 0.04	& 0.9 & $1.8 \pm 0.2 $		& $25 \pm 1$ 		& $7.6 \pm 0.1$ & 71.5(81) & 0.76626\\
081221 	& 196.3 	& 0.2	& 0.9 & $0.40 \pm 0.02$ 	& $97 \pm 3$ 		& $6.8 \pm 0.2$ & 268.7(259) & 0.32575\\
090205 	& 321.5 	& 0.02	& 0.9 & $0.5 \pm 0.1$ 		& $61 \pm 5$ 		& $11.0 \pm 0.6$ & 18.6(31) & 0.96039\\
090418A & 101.7 	& 0.01	& 0.9 & $0.7 \pm 0.1$ 		& $31 \pm 1$ 		& $4.1 \pm 0.1$ & 7.8(111) & 1.0\\
091029 	& 471.9 	& 0.04	& 0.9 & $0.11 \pm 0.01$ 	& $26 \pm 1$ 		& $7.1 \pm 0.4$ & 120.3(130) & 0.71638\\
100219A & 464.5 	& 0.1	& 0.9 & $1.2 \pm 0.3$	 	& $9\pm1$ 			& $2.7 \pm 0.1$ & 44.2(26) & 0.014207\\
100906A & ... & ... & ... & ... & ... & ... & ... & ... \\
110213A & 148.4 	& 0.02	& 0.9 & $0.89 \pm 0.03$	 	& $45.1 \pm 0.9$ 	& $7.0 \pm 0.1$ & 243.8(227) & 0.21048\\
110422A & 85.8 		& 0.2	& 0.9 & $0.50 \pm 0.03$ 	& $62 \pm 2$ 		& $8.0 \pm 0.1$ & 173.4(267) & 1.0\\
111008A & 241.8 	& 0.1	& 0.9 & $0.29 \pm 0.03$ 		& $16.4 \pm 0.6$ 		& $3.1 \pm 0.1$ & 71.9(137) & 1.0\\
120118B & 257.4 	& 0.004 & 0.9 & $0.2 \pm 0.1$ 		& $51 \pm 4$ 		& $9.1 \pm 1.5$ & 8.4(34) & 1.0\\
120327A & 163.2 	& 0.3	& 0.9 & $0.57 \pm 0.04$ 		& $38 \pm 2$ 		& $4.6 \pm 0.1$ & 32.1(59) & 0.99832\\
120404A & 438.3 	& 0.08	& 0.9 & $0.6 \pm 0.1$ 		& $66 \pm 4$ 		& $8.2 \pm 0.9$ & 17.5(36) & 0.99604\\
120521C & 650.5 	& 0.06	& 0.9 & $0.4 \pm 0.1 $		& $33 \pm 6$ 		& $8.5 \pm 0.8$ & 0.5(6) & 0.99764\\
120811C & 216.4 	& 0.1	& 0.9 & $0.40 \pm 0.07$ 		& $38 \pm 3$ 		& $5.1 \pm 0.2$ & 32.2(46) & 0.93796\\
120907A & 90.3 		& 0.04	& 0.9 & $0.11 \pm 0.02$ 		& $104 \pm 6$ 		& $10.4 \pm 0.8$ & 66.3(87) & 0.95166\\
120922A & 657.8 	& 0.05	& 0.9 & $0.40 \pm 0.05$ 		& $37 \pm 4$ 		& $10.7 \pm 0.4$ & 21.8(70) 	& 0.999999\\
121128A & 127.1 	& 0.3	& 0.9 & $0.48 \pm 0.04$ 		& $68 \pm 5$ 		& $5.3 \pm 0.1$ & 67.0(90) & 0.966758\\
131105A & 327.8 	& 0.04	& 0.9 & $0.12 \pm 0.02$ 		& $57 \pm 4$ 		& $11 \pm 1$ & 16.3(50) & 1.0\\\
131117A & 312.3 	& 0.1	& 0.9 & $0.21 \pm 0.03$ 	& $47 \pm 9$ 		& $13 \pm 1$ & 0.36(11) & 1.0\\\
140206A & 411.38 	& 1.0 	& 0.9 & $0.48 \pm 0.02$		& $9.3 \pm 0.2$ 	& $1.59 \pm 0.02$ & 315.6(442) & 1.0\\
140506A & 573.5 	& 0.4	& 0.9 & $0.17 \pm 0.02$ 		& $15 \pm 1$ 	& $5.2 \pm 0.2$ & 88.8(140) & 0.999763\\
140512A & 326.9 	& 0.2	& 0.9 & $0.51 \pm 0.01$ 		& $28 \pm 1$ 	& $5.9 \pm 0.1$ & 325.8(363) & 0.920167\\
\hline
051221A	& 1868.9	& 0.002	&	0.9	& $0.25\pm0.05$ & $104\pm9$ 	& $44\pm3$ 		& 	62.1(47)	& 0.06815\\
060614A	& 1630.5	& 3.3e-06&	0.9	& $1.3\pm0.1$ 	& $156\pm	5$ 	& $111\pm1$ 	& 	151.5(140)	& 0.23827\\
060912A	& 143.5		& 0.01	&	0.9	& $0.13\pm0.03$ & $435\pm	65$ & $35\pm7$ 		& 	43.0(33)	& 0.11397\\
061201	& 9.8		& 0.02	&	0.9	& $0.9\pm0.1$ 	& $206\pm	15$ & $22\pm0.7$	& 	29.8(24)	& 0.19257\\
070714B	& 572.1		& 0.01	&	0.9 & $0.8\pm0.2$	& $369\pm40$	& $35\pm12$		&	39.0(20)	& 0.00629\\
070809	& 497.9		& 0.002	&	0.9	& $0.3\pm0.1$ 	& $58\pm1	1$ 	& $20\pm2$ 		& 	16.4(13)	& 0.23016\\
070810A	& 167.8		& 0.01	&	0.9	& $0.4\pm0.1$ 	& $178\pm	17$ & $17\pm2$ 		& 	7.0(29)		& 0.99999\\
090426	& 1.5		& 0.01	&	0.9	& $0.10\pm0.05$ & $212\pm	33$ & $10\pm2$ 		& 	13.8(27)	& 0.982628\\
090510	& 1.4		& 0.02	&	0.9	& $1.0\pm0.1$ 	& $116\pm	5$ 	& $8\pm0.1$ 	& 	83.4(70)	& 0.13127\\
100724A	& 55.7		& 0.01	&	0.9	& $0.4\pm0.1$ 	& $255 \pm 38$ 	& $34\pm2$ 		& 	11.9(15)	& 0.68291\\
130603B & ... & ... & ... & ... & ... & ... & ... & ... \\
\hline
\end{tabular}
\end{table}

\bibliographystyle{aasjournal}
\bibliography{biblioReview5} 

\end{document}